\documentclass[aps,superscriptaddress,reprint]{revtex4-2}

\usepackage[utf8]{inputenc}
\usepackage{xcolor}
\usepackage{pst-node}
\usepackage{graphicx}
\usepackage{amsmath, amssymb, amsthm}
\usepackage{mathtools}
\usepackage{hyperref}
\usepackage[shortlabels]{enumitem}
\usepackage{dsfont}
\usepackage{amsthm}
\usepackage{amssymb}
\usepackage{amstext}
\usepackage{braket}
\usepackage{amsfonts}
\usepackage{nicefrac}
\usepackage{extarrows} 
\usepackage{graphicx}
\usepackage{relsize}
\usepackage{soul}
\usepackage{xfrac} 

\newcommand{\skiptext}[1]{}

\begin{document}

\title{Hunting for quantum advantage in electronic structure calculations\\ is a highly non-trivial task}

\author{\"Ors Legeza}
\email{legeza.ors@wigner.hu}
\affiliation{%
Strongly Correlated Systems Lend\"ulet Research Group,
Wigner Research Centre for Physics, H-1525, Budapest, Hungary
}%
\affiliation{Dynaflex LTD, Zr{\'i}nyi u 7, 1028 Budapest, Hungary}
\affiliation{
Institute for Advanced Study, Technical University of Munich, Germany, Lichtenbergstra{\ss}e 2a, 85748 Garching, Germany
}
\affiliation{Parmenides Stiftung, Hindenburgstr. 15, 82343, P{\"o}cking, Germany}

\author{Andor Menczer}
\email{menczer.andor@wigner.hu}
\affiliation{%
Strongly Correlated Systems Lend\"ulet Research Group,
Wigner Research Centre for Physics, H-1525, Budapest, Hungary
}%
\affiliation{%
E{\"o}tv{\"o}s Lor{\'a}nd University, P\'azm\'any P\'eter S\'et\'any 1/C, 1117 Budapest, Hungary
}%

\author{Mikl\'os Antal Werner}
\email{werner.miklos@wigner.hun-ren.hu}
\affiliation{%
Strongly Correlated Systems Lend\"ulet Research Group,
Wigner Research Centre for Physics, H-1525, Budapest, Hungary
}%

\author{Sotiris S. Xantheas}
\email{sotiris.xantheas@pnnl.gov and xantheas@uw.edu}
\affiliation{%
Advanced Computing, Mathematics, and Data Division, Pacific Northwest National Laboratory, Richland, Washington 99354, USA}%
\affiliation{%
Department of Chemistry, University of Washington, Seattle, WA 98195, USA}

\author{Frank Neese$^\ast$}
\email{neese@kofo.mpg.de}
\affiliation{
Max-Planck Institut für Kohlenforschung, Kaiser-Wilhelm-Platz 1,\\ D-45470 M{\"u}lheim an der Ruhr, Germany}

\author{Martin Ganahl}
\email{martin.ganahl@sandboxaq.com}
\affiliation{SandboxAQ, Palo Alto, California, USA}%

\author{Cole Brower}
\email{cbrower@nvidia.com}
\affiliation{%
NVIDIA, 2788 San Tomas Expressway, Santa Clara, CA 95051
}%

\author{Samuel Rodriguez Bernabeu}
\email{srodriguezbe@nvidia.com}
\affiliation{%
NVIDIA, 2788 San Tomas Expressway, Santa Clara, CA 95051
}%

\author{Jeff Hammond}
\email{jeffpapers@nvidia.com}
\affiliation{%
NVIDIA Helsinki Oy, Porkkalankatu 1, 00180 Helsinki
}%

\author{John Gunnels}
\email{jgunnels@nvidia.com}
\affiliation{%
NVIDIA, 2788 San Tomas Expressway, Santa Clara, CA 95051
}%

\date{\today}

\begin{abstract}
In light of major developments over the past decades in both quantum computing and simulations on classical hardware, it is a serious challenge to identify a real-world problem where quantum advantage is expected to appear.
In quantum chemistry, electronic structure calculations of strongly correlated, i.e. multi-reference problems, are often argued to fall into such category because of their intractability with standard methods based on mean-field theory.
Therefore, providing state-of-the-art benchmark data by classical algorithms is necessary to make a decisive conclusion when such competing development directions are compared. 
In this work, we report cutting-edge performance results together with high accuracy ground state energy for the Fe$_4$S$_4$ molecular cluster
on a CAS(54,36) model space, a problem that has been
included quite recently among the list of systems in the {\it Quantum Advantage Tracker} webpage maintained by IBM and RIKEN.
Pushing the limits even further, we also present 
CAS-SCF based orbital optimizations for unprecedented CAS sizes of
up to 89 electrons in 102 orbitals [CAS(89,102)] for the 
Fe$_5$S$_{12}$H$_4^{5-}$
molecular system comprising 
twenty five open shell orbitals in its sextet ground state and an active spaces size
of 331 electrons in 451 orbitals.
We have achieved our results
via mixed-precision spin-adapted \textit{ab initio} Density Matrix Renormalization Group (DMRG) electronic structure calculations
interfaced with the ORCA program package and
utilizing the NVIDIA Blackwell graphics processing unit (GPU) platform.
We argue that DMRG benchmark data should be taken as a classical reference when quantum advantage is reported.
In addition, full exploitation of classical hardware should also be considered since even the most advanced DMRG implementations are still in a premature stage regarding utilization of all the benefits of GPU technology.
\end{abstract}

\maketitle

{\it Introduction:} 
Over the past decades, major developments in both classical and quantum computation have brought the question of the realization of quantum advantage in real-life problems into the focus of modern science~\cite{Lee-2023,Alexeev-2024, Lanes-2025, Eisert-2025, Altman-2021, Bharti-2022, Cerezo-2021, Preskill-2018, Arute-2019, Wu-2021, Madsen-2022, Zhong-2020, Kim-2023}.
However, scalable electronic structure simulation workflows have only recently been demonstrated on quantum hardware~\cite{Yamamoto-2025} even for basic systems. While recent experiments have reached regimes where classical simulation of quantum dynamics becomes extremely challenging~\cite{Haghshenas-2025, Abanin-2025} it remains difficult to confidently identify for which problems of practical interest can quantum computers reliably outperform classical computers in the near future~\cite{Babbush-2025}.
Nevertheless, recent advances in  
quantum chemistry~\cite{Eddins-2022,Fan-2022,Liepuoniute-2024,Santagati-2024, Alexeev-2024, RobledoMoreno-2025,Quantinuum-2025,Smith-2025,Jornada-2025}, condensed matter systems ~\cite{Kanno-2025,Duriez-2025,Kan-2025}, high energy physics~\cite{DiMeglio-2024},  quantum dynamics~\cite{Google-2025, Fischer-2026}, quantum thermal simulation~\cite{Chen-2025} optimization problems~\cite{Abbas-2024, Koch-2025}, and even in simulation of classical systems~\cite{Babbush-2023, Li-2025} or life sciences~\cite{Basu-2023} 
predict great opportunities in the future.

While fault-tolerant quantum computing is a promising direction~\cite{Gottesman-2022,Omanakuttan-2024,Litinski-2025}, the underlying hardware technology is still in a development stage.  
In the case of noisy quantum simulations, the reduction of noise requires various error mitigation techniques and serious post-processing, whose use strongly limits the maximal accessible complexity of quantum circuits ~\cite{Temme-2017, Endo-2018, Kandala-2019, Takagi-2022, Cai-2023, vandenBerg-2023,Zimboras-2025}.
In contrast, simulation of quantum systems on classical hardware is free of such artifacts; however, the required computational resources scale exponentially with the amount of entanglement and correlations encoded in the underlying quantum many-body wave function~\cite{Schollwock-2011,Chan-2012,Szalay-2015a,Orus-2019,Baiardi-2020,Cheng-2022,Verstraete-2023}.
A promising class of ansatz for an efficient data-sparse representation of the wave functions can be realized via low-order tensor-product factorization, also known as the tensor network state (TNS) methods~\cite{White-1993, White-1996, Fannes-1989, Ostlund-1995, Rommer-1997, Verstraete-2004a, Vidal-2007, Schollwock-2011,Verstraete-2023}. At the current level of development of quantum hardware, TNS benchmark simulations together with quantum computations on real quantum chips are mandatory to cross-validate results, and they also provide great impetus and shed new light on our understanding of various hard problems in modern physics~\cite{McCaskey-2018, Fan-2022, Gray-2021, Kanno-2025, RobledoMoreno-2025, Smith-2025, Cobos-2025}. 

While parallelism is intrinsically achieved via quantum simulations, today modern high performance computing (HPC) via Graphics Processing Units (GPUs)~\cite{a100,nvidia-gh200,blackwell,mi300}  also offers unprecedented computational power via massive parallelization
\cite{Hager-2004,Stoudenmire-2013,Nemes-2014,Ganahl-2019, Milsted-2019,Brabec-2021,Zhai-2021,Gray-2021,Unfried-2023,Ganahl-2023,Menczer-2023a,Menczer-2023b,Menczer-2024a,Menczer-2024b,Xiang-2024,Menczer-2024c,Werner-2025}.
This is further boosted by highly improved data communication technologies~\cite{nvl72}, allowing several hundreds of thousands of computing units to work together on a single task or even on a collection of a large number of linear algebra based operations.

In this work, we report cutting-edge TNS performance results together with state-of-the-art accuracy ground state energy for the Fe$_4$S$_4$ molecular cluster~\cite{Sharma-2014c,Jiang-2024}. This system has been identified as a very difficult problem  where quantum advantage could be realized, and thus has been included recently among systems listed in the {\it Quantum Advantage Tracker} webpage maintained by IBM and RIKEN~\cite{ibm-web}.
In addition, we introduce new benchmark systems via CAS-SCF based orbital optimizations~\cite{Zgid-2008b} for unprecedented CAS sizes of up to 89 electrons in 102 orbitals [CAS(89,102)] for the
Fe$_5$S$_{12}$H$_4^{5-}$
molecular system comprising twenty five open shell orbitals in its sextet ground state. 
We have, thus performed
mixed-precision spin adapted ab initio Density Matrix Renormalization Group (DMRG) electronic
structure calculations utilizing state-of-the-art NVIDIA Blackwell technology~\cite{blackwell}.

{\it Method:} 
As the foundational Tensor Network State (TNS) algorithm, the Density Matrix Renormalization Group (DMRG) serves as a variational optimization method designed to find the ground state of a model Hamiltonian, ($\cal{H}$), within the Matrix Product State (MPS) manifold ~\cite{Schollwock-2011}. For a quantum chemical system described in terms of $N$ spin-orbitals ~\cite{White-1999} where each site is defined by the local basis $\ket{i_n} = \{\ket{0}, \ket{\uparrow}, \ket{\downarrow}, \ket{\uparrow\downarrow}\}$, the global wave function is represented as
\begin{equation}
  \ket{\Psi_{MPS}}  = \sum_{\{i_k\}} \sum_{\{\alpha_p\}}[A_1]_{1\alpha_1}^{i_1} [A_2]_{\alpha_1\alpha_2}^{i_2} \dots [A_{N}]_{\alpha_{N-1}1}^{i_{N}} \ket{i_1\dots i_k} \; .
\label{eq:mps}
\end{equation}
In this formulation, $A^{i_n}_{\alpha_{n-1}\alpha_n}$ are order-3 tensors of dimension $(D_{n-1},4,D_n)$, with the exception of the boundaries (the first and last orbitals), which are represented by order-2 tensors.
The numerical precision of the ansatz is governed by the matrix ranks, $D$, commonly referred to as the bond dimension; increasing $D$ systematically improves the approximation. While the exact solution is only recovered when the bond dimension scales exponentially -- specifically $D_n\sim4^n$ for $1\le n \le N/2$ and  $D_n\sim4^{N-n}$ for $N/2\leq n\leq N$ -- practical applications rely on truncating these ranks to maintain computational feasibility. Under these approximations, the memory and computational requirements scale as $O(N^2D^2)$ and $O(N^4D^3)$, respectively.

Since DMRG is a local optimization protocol, the overall numerical error is determined by two main sources~\cite{White-1992a,Legeza-1996,Schollwock-2005}: {\em the truncation error} which is related to the discarded Schmidt-weights and thus to the bond dimension and {\em the environment error} which describes the quality of embedding when the full system is partitioned into two blocks. By using a large enough number of sweeps, and for larger $D$ values, the environment error can be eliminated, therefore, the overall accuracy is determined in leading order by the truncation error.  

{\it Numerical procedure:}
As the main purpose of the current work is to provide high accuracy classical benchmark data for quantum computing, we have used a fixed number of $SU(2)$ multiplets, $D_{SU(2)}\in\{256,512,1024,2048,4096,8192,10240,12280\}$ and increased it systematically.
Note that, alternatively, one could fix the truncation error and vary bond dimension dynamically~\cite{Legeza-2003a}, and utilize also fermionic mode optimization~\cite{Krumnow-2016,Menczer-2024a,Friesecke-2024}, but the currently applied protocol offers a more straightforward procedure for data reproduction. For each $D_{SU(2)}$ value the sweeping procedure was terminated once the energy difference between three subsequent sweeps dropped below $\varepsilon_{\rm sweep}=10^{-5}$
and we set the residual error of the iterative L\'anczos diagonalization error to $\varepsilon_{\rm Lanczos}=10^{-6}$. 
We found that for smaller $D$ values some 15-20 sweeps were used while for larger $D$ usually 7-9 sweeps were sufficient to fulfill our error criteria. The chain-like DMRG topology was preoptimized using quantum  information based protocols~\cite{Barcza-2011,Szalay-2015a}.

{\it Hardware specific implementations:} 
On the NVIDIA Blackwell architecture, the
cuBLAS library introduces versatile interfaces for testing and deploying FP64 arithmetic emulation. Users can control this behavior via environment variables or dedicated APIs to toggle emulation, specify mantissa bit depth, or allow the library to dynamically optimize these values. The cuBLAS library follows a specific hierarchy for precision:
\textit{Default Behavior} -- In the absence of environment variables, the system defaults to native FP64. \textit{Performant Mode} -- If emulation is enabled without a manual mantissa bit count, the library automatically calculates the bits required to maintain native FP64 accuracy. \textit{Eager Mode} -- Eager mode enforces emulation with specified mantissa bit count
regardless of the specific performance profile of the operation. It is primarily effective for compute-bound tasks. 
By representing the underlying DMRG matrix and tensor algebra as a collection of fixed-point "slices," we can systematically bridge the gap between half-precision and double-precision performance. In this work, we evaluate three primary configurations: native FP64 double-precision, performant mode, utilizing dynamic mantissa bit settings, and eager mode enforcing fixed $47$-bit mantissa ($\kappa=6$ slices),
providing a high-fidelity approximation of standard double-precision arithmetic. In all cases the emulation of
FP64 arithmetic is based on the Ozaki scheme~\cite{Ootomo-2024a,Uchino-2025,Dongarra-2024,Brower-2025}.

\begin{figure}
    \centering    
    \includegraphics[width=0.48\textwidth]{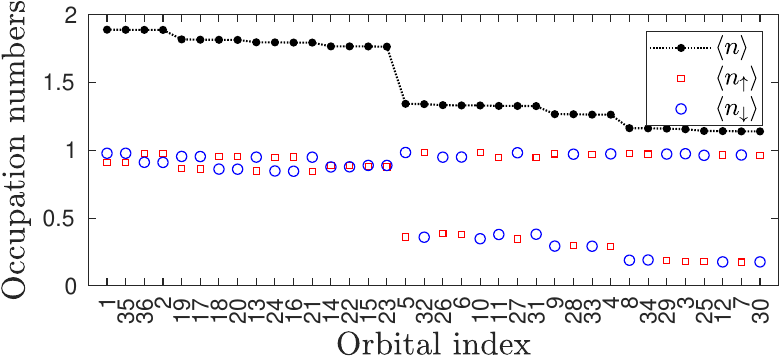}
    \caption{
    Orbital occupation number profile obtained for the Fe$_4$S$_4$ cluster in a CAS(54,36) model space using DMRG with $D=8192$ and 20 sweeps, and the basis given in Ref.~\cite{Sharma-2014c}. 
    $\langle n_\uparrow\rangle$ and $\langle n_\downarrow\rangle$ stand for occupation number of up- and down-spin electrons, respectively, while $\langle n\rangle=\langle n_\uparrow + n_\downarrow\rangle$. This latter quantity is also reproduced directly using $D_{SU(2)}=4096$ multiplets. 
    Labels in the x-axis indicate the corresponding orbital index.
    }
    \label{fig:occup_gs}
\end{figure}

{\it Model system:}
Iron-sulfur clusters have long been used as benchmark systems for numerical method developments~\cite{Sharma-2014c,Jiang-2024,MejutoZaera-2022,Tzeli-2024,Legeza-2025a,zhai2026classicalsolutionfemocofactormodel}. In fact, the minimal CAS(54,36) model space, considering the correlation of 54 electrons on 36 orbitals of Fe$_4$S$_4$, introduced in Ref.~\cite{Sharma-2014c}, has been included recently among the list of hard problems for classical computation in a webpage maintained by IBM and RIKEN~\cite{ibm-web}. The difficulty stems from the existence of many open shell orbitals where orbital occupancy is neither close to two or zero. This multi-reference nature is demonstrated clearly in Fig.~\ref{fig:occup_gs}  
as a result of our DMRG calculation using $D=8192$ and 20 sweeps.

\begin{figure}
    \centering        \includegraphics[width=0.48\textwidth]{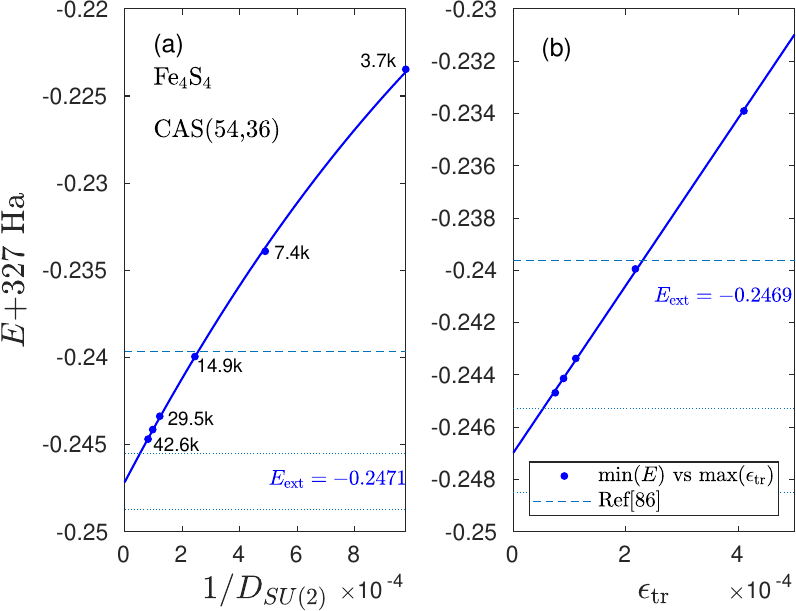}
    \caption{The ground state energy as a function of inverse DMRG bond dimension, $D_{SU(2)}\in\{1024,\ldots,12288\}$ for the Fe$_4$S$_4$ molecule in a CAS(54,36) model space ~\cite{Sharma-2014c} (left panel) and as a function of truncation error $\varepsilon_{\rm tr}$
    (right panel) obtained on a DGX B200 system via performant mode.
    Solid lines stand for second-order polynomial fits. The corresponding $U(1)$ bond dimension values are indicated by numbers next to the data points. Dashed line stands for reference energy obtained earlier with $D_{SU(2)}=4000$~\cite{Sharma-2014c} while dotted lines indicate 1.6 milliHa chemical accuracy with respect to the extrapolated ground state energy, $E_{\rm ext}$. 
    }
    \label{fig:egs_fe4s4}
\end{figure}

{\it Numerical benchmark:}
In order to provide high accuracy classical benchmark data we have performed large scale spin adapted fixed-$D$ DMRG calculations utilizing native FP64 double precision and carried out extrapolation to the truncation-free limit using two different approaches.
In Fig.~\ref{fig:egs_fe4s4}(a), the shifted ground state energy is shown as a function of inverse bond dimension for $D_{SU(2)}\ge1024$, where the environment error is small enough, i.e., the overall error is determined mainly by the truncation error. 
The solid line is a second-order polynomial fit leading to $E_{\rm ext}=-327.2471$ Ha in the $D\rightarrow\infty$ limit. 
In the second approach, in Fig~\ref{fig:egs_fe4s4}(b), the minimum energy as a function of the maximum truncation error obtained in the last DMRG sweep is plotted. Here, the extrapolation resulted in
$E_{\rm ext}=-327.2469$, thus the difference between the two methods is $\delta E=0.2$ milliHa. The dashed line in both panels represents the reference energy reported earlier in Ref.~\cite{Sharma-2014c} using $D_{SU(2)}=4000$, while the dotted lines show the $\pm 1.6$ milliHa chemical accuracy with respect to the extrapolated ground state energy, $E_{\rm ext}$. Numbers next to the data points indicate the corresponding maximum $U(1)$ bond dimension values.
Since our result is much more accurate than those reported earlier, we claim that such a reference should be used when quantum simulations are benchmarked.
Regarding performance and computational time, we measured a maximum of around 220 TFLOPS for $D_{SU(2)}=12288$ 
and the calculation resulting in $E(D)=-327.24466$ 
took about 12.6 hours on a DGX B200 node.  
While completing the current work, we became aware of a recent study via large-scale DMRG on the FeMoco cluster~\cite{zhai2026classicalsolutionfemocofactormodel} also including benchmark results on the Fe$_4$S$_4$ cluster. In that work, both the extrapolated value of $-327.245273$ Ha, and the calculated bare DMRG ground state energy of
$-327.244001$
with $D_{SU(2)}=18000$ slightly lie higher than our results.
This may account for our improved SU(2) aware Dynamically Extended Active Space (SU(2)-DEAS) procedure~\cite{Legeza-2003b, Werner-2026a}, which has a significant effect on the convergence of DMRG for systems having a large number of open shell orbitals.

{\it Mixed precision and hardware:} 
Repeating our analysis, but utilizing the Ozaki scheme
for emulating FP64 arithmetic through the use of fixed-point compute resources, we obtained a very similar result.
We found that by enforcing mantissa bit setting to 47 ($\kappa=6$ INT8 slices) the largest absolute error measured with respect to the native FP64 data set was less than $10^{-4}$. Since this error is almost invisible in Fig.~\ref{fig:egs_fe4s4} we conclude that the quality of our extrapolation scheme is not affected within the error margin set by the chemical accuracy. Therefore, development directions utilizing mixed precision arithmetic in Blackwell technology do not pose any negative impact on the scientific application of the new hardware.

{\it New benchmark system for quantum advantage tracker:}
In light of our high accuracy results discussed above, we pose a more challenging problem that could be considered as a new classical benchmark for quantum computation. Therefore, we performed 
an efficient orbital optimization procedure by combining our highly GPU-accelerated, spin-adapted DMRG method~\cite{Menczer-2023a,Menczer-2023b,Menczer-2024c} with the complete active space self-consistent field (CAS-SCF) approach for quantum chemistry implemented in the ORCA program package~\cite{neese_orca_2020} 
on the Fe$_k$S$_{2k-2}$(SH)$_4^{k-}$ cluster for $k=4$ and 5,
introduced recently~\cite{Legeza-2025a}.
In this system, the local spins located in the iron d-based molecular orbitals are all aligned parallel, leading to local S(Fe)=5/2 fragments. The different local fragments couple antiferromagnetically, leading to overall low spin states with total spin zero for $k=4$ and with total spin 5/2 for $k=5$, e.g. for the dimeric species to an antiferromagnetic singlet with twenty
and a sextet with twenty-five 
unpaired electrons, respectively.
In the DMRG active space, we employed localized orbitals that are properly aligned to belong to the iron ions, and it contained the 
3d, 4s, 4p, and 4d orbitals of iron, 
and the 3p orbitals of the sulfur atoms, which are marked as the 'B' set of orbitals in Ref.~\onlinecite{Legeza-2025a}.
By taking $k=4$, we fully correlate 72 electrons in 82 orbitals on a [CAS(72,82)] model space, while for $k=5$ we fully correlate 89 electrons in 102 orbitals on a [CAS(89, 102] model space.
Note that the full orbital space corresponds to the correlation of $N_{\rm e}=272$ electrons in $N_{\rm o}=372$ orbitals for $k=4$ and for $N_{\rm e}=331$ electrons in $N_{\rm o}=451$ orbitals for $k=5$. 

\begin{figure}
    \centering    
    \includegraphics[width=0.48\textwidth]{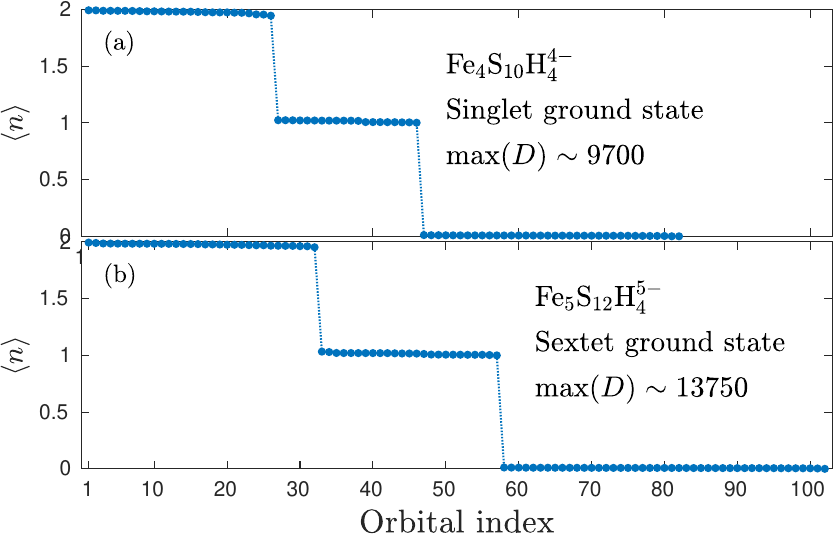}
    \caption{
    Orbital occupation number profile, $\langle n \rangle$, obtained for the DMRG-SCF optimized basis via $D_{SU(2)}=2048$ for (a) 
    the singlet ground state of Fe$_4$S$_{10}$H$_4^{4-}$ cluster in a CAS(72,82) model space 
    and for (b) the
    sextet ground state of the Fe$_5$S$_{12}$H$_4^{5-}$ cluster in a CAS(82,102) model space. 
    The twenty and twenty-five open shell orbitals correspond to the four and five Fe atoms, respectively, while
    $\max(D)$ indicates the corresponding maximum $U(1)$ bond dimension.
    }
    \label{fig:occup_fe6s14}
\end{figure}

\begin{figure}
    \centering    
    \includegraphics[width=0.48\textwidth]{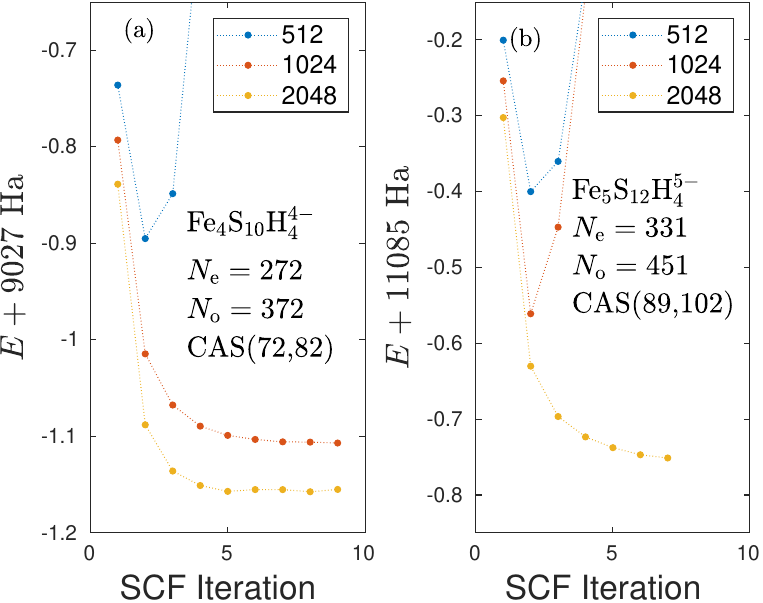}
    \caption{
    Convergence of DMRG-SCF for the (a) singlet ground state of Fe$_4$S$_{10}$H$_4^{4-}$ cluster in a CAS(72,82) model space using various $D_{SU(2)}\in\{512,1024,2408\}$, $\varepsilon_{\rm sweep}=10^{-5}$, 
    $\varepsilon_{\rm Lanczos}=10^{-6}$
    and orca ``tight" error setting. (b) The same but for the sextet ground state of
    Fe$_5$S$_{12}$H$_4^{5-}$ in a CAS(89,102) model space. 
    }
    \label{fig:dmrgscf_fe6s14}
\end{figure}

Our results are summarized in Figs~\ref{fig:occup_fe6s14} and ~\ref{fig:dmrgscf_fe6s14}.
The obtained occupation number profile via our DMRG-SCF procedure
using $D_{SU(2)}=2048$, 
reflecting the twenty and twenty-five open shell orbitals for $k=4$ and $k=5$, are 
shown in Fig.~\ref{fig:occup_fe6s14}. The large number of open shell orbitals poses a significant challenge: there are many states with the same occupation number profile but different spin configurations. Therefore, even when $SU(2)$ symmetry is exploited, the number of relevant configuration state functions (CSF's) is very large, leading to a high required bond dimension ($\max D\simeq 9700$ and 13750 for $k=4$ and 5, respectively).
In Fig.~\ref{fig:dmrgscf_fe6s14}
the convergence profile of the DMRG-SCF protocol is presented for various $D_{SU(2)}$ bond dimension values.
Here we remark that convergence issues reported previously for $2\le k\le3$ in Ref.~\cite{Legeza-2025a} have been resolved in our current DMRG implementation. 
The detailed convergence analysis for $2\le k\le6$ together with technical details of the implementation will be part of a subsequent publication. 
To achieve stable convergence, the absolute error in the DMRG calculation must be kept below the error setting of the gradient descent method in the SCF step~\cite{Legeza-2025a}, which requires a relatively larger bond dimension value, i.e., for the large active spaces considered here, $D_{SU(2)}\ge 1024$ and $D_{SU(2)}\ge 2048$ is mandatory for $k=4$ and 5, respectively. 
Alternatively, the truncation error can be fixed to guarantee stable convergence as has been demonstrated in Ref.~\cite{Legeza-2025a}. 
The importance of treating dynamic correlations properly via the SCF procedure is clear as it results in an energy reduction of the order of $10^{-1}$ Ha.
Moreover, performing also a local spin analysis ~\cite{Clark-2001} by partitioning the orbitals of the Fe atoms and the orbitals of the remaining S and H atoms into separate fragments, the expectation value of the total spin operator of each Fe fragment turned out to be $2.471(3)$. 
The expectation value of the spin-spin correlation function between neighboring Fe fragments is negative, once the converged orbital set is used. These results confirm with high accuracy the expected antiferromagnetic configuration of the coupled spin-5/2 states of the Fe atoms in the ground state.

In order to study the effect of
mixed-precision arithmetic, we have repeated the DMRG-SCF workflow by setting mantissa bits to 47 in the DMRG algorithm. We found stable convergence again and an absolute error of $10^{-5}-10^{-6}$ with respect to the native FP64 data set. Therefore, even DMRG-SCF can be performed safely via reduced precision on the Blackwell supercomputer, opening new research directions on the new hardware.

{\it Unexploited power and future perspectives:}
Despite presenting the largest classical TNS simulations to date, state-of-the-art hardware technology offers tremendous additional possibilities for future developments that have not been utilized in the current work. 
First, the data transfer I/O time between host and device, i.e., transferring data between CPU and GPU, can be reduced significantly by utilizing low latency/high bandwidth coherency systems
such as NVIDIA Grace 
superchips~\cite{nvidia-gh200,blackwell} or AMD MI300~\cite{mi300} platforms. This will reduce and shift the performance decline observed for large bond dimension~\cite{Brower-2025} to much larger $D$ values.   
Moreover, performance can be further boosted via an efficient multi-GPU multi-node (MGMN) DMRG. 
Our pilot work in this direction, based on MPI protocols reaching PetaFLOPS performance~\cite{Menczer-2023c}, however, does not show reasonable scaling due to the limited bandwidth of InfiniBand (note that double InfiniBand has 25 GB/s bandwidth, see also Ref.\cite{Menczer-2024c}). In contrast to this, NVIDIA GB200 NVL72 utilizes fifth-generation NVLink, which connects up to 576 GPUs in a single NVLink domain with over 1 PB/s total bandwidth and 240 TB of fast memory~\cite{nvl72}. This, together with a revolutionary 1.8 TB/s of bidirectional throughput per GPU
has the potential to push the performance of DMRG well into the PetaFLOPS regime.
In addition, further improvements in mixed-precision arithmetic~\cite{Ootomo-2024a,Uchino-2025,Dongarra-2024,Brower-2025} could also accelerate DMRG calculations even more.
Finally, recent hardware such as B300~\cite{nvidia-gb300} offers even more GPU memory, 
which is very beneficial for ab initio DMRG dealing with large data sets.

{\it Conclusion:} While parallelism, in some sense, is intrinsically encoded in quantum hardware, the development of classical hardware and related massively parallel numerical algorithms also provides unprecedented computational power to simulate quantum systems efficiently. Most such algorithms are, however, still in a premature stage regarding the utilization of all the benefits offered by state-of-the-art classical hardware.
In this work, we have provided a classical benchmark for a hard problem identified recently by IBM and RIKEN, where quantum advantage is expected to be gained. 
Besides high accuracy ground state energy and a performance analysis obtained via a recent NVIDIA Blackwell supercomputer, we pushed classical limits even further via a CAS-SCF-based orbital optimization on an active space size of several hundreds of electrons on several hundreds of orbitals.
Utilization of further, as yet unexploited, technologies is also discussed. These have the potential to shift the performance of DMRG towards the exascale limit, opening new research directions in electronic structure calculations, materials science, and beyond~\cite{Legeza-2025a}.
Since the presented DMRG-SCF framework is variational, future simulations with larger bond dimensions and on larger active spaces
can be carried out periodically, improving systematically extrapolations and bare classical references.
Therefore, demonstration of quantum advantage in electronic structure calculations will be a delicate issue and will require rigorous benchmarks against state-of-the-art classical simulations.

{\it Acknowledgments:} This work has been supported 
by the Center for Scalable and Predictive methods for Excitation and Correlated phenomena (SPEC), funded as part of the Computational Chemical Sciences, FWP 70942, by the U.S. Department of Energy (DOE), Office of Science, Office of Basic Energy Sciences, Division of Chemical Sciences, Geosciences, and Biosciences at Pacific Northwest National Laboratory. This work has also been supported by the Hungarian
National Research, Development and Innovation Office
(NKFIH) through Grant No. STARTING 152628.
\"O.L. acknowledges financial support
by the Hans Fischer Senior Fellowship programme funded by the Technical University
of Munich – Institute for Advanced Study.
We also thank the computational support provided during the initial phase of the project from the Wigner Scientific Computing Laboratory (WSCLAB) and the national supercomputer HPE Apollo Hawk at the High Performance Computing Center Stuttgart (HLRS) under the grant number MPTNS/44246.




\providecommand{\latin}[1]{#1}
\makeatletter
\providecommand{\doi}
  {\begingroup\let\do\@makeother\dospecials
  \catcode`\{=1 \catcode`\}=2 \doi@aux}
\providecommand{\doi@aux}[1]{\endgroup\texttt{#1}}
\makeatother
\providecommand*\mcitethebibliography{\thebibliography}
\csname @ifundefined\endcsname{endmcitethebibliography}
  {\let\endmcitethebibliography\endthebibliography}{}

\end{document}